# Theoretical Proposal for Determining Angular Momentum Compensation in Ferrimagnets

Zhifeng Zhu[a)] Xuanyao Fong, and Gengchiau Liang[b)]
Department of Electrical and Computer Engineering, National University of Singapore, Singapore

This work demonstrates that the magnetization and angular momentum compensation temperature ($T_{MC}$ and $T_{AMC}$) in ferrimagnets (FiM) can be unambiguously determined by performing two sets of temperature dependent current switching, with the symmetry reverses at $T_{MC}$ and $T_{AMC}$, respectively. A theoretical model based on the modified Landau-Lifshitz-Bloch equation is developed to systematically study the spin torque effect under different temperatures, and numerical simulations are performed to corroborate our proposal. Furthermore, we demonstrate that the recently reported linear relation between $T_{AMC}$ and $T_{MC}$ can be explained using the Curie-Weiss theory.

Magnetization dynamics of ferrimagnets (FiM) driven by spin-orbit torque (SOT) has attracted considerable attention, especially in material with antiferromagnetic coupled transition-metal (TM) and rare-earth (RE) alloy (e.g. $Gd_X(FeCo)_{1-X}$ or $Co_{1-X}Tb_X$) [1-7]. The magnetization in FiM can be tuned through temperature ($T$) [2, 3] or material composition ($X$) [4-6], resulting in the compensation point ($T_{MC}$ or $X_{MC}$) with zero net magnetization ($m_{net}$). The different g-factors of TM and RE induce another compensation ($T_{AMC}$ or $X_{AMC}$) where the net angular momentum ($S_{net}$) vanishes [8]. Furthermore, FiM has faster dynamics than ferromagnet (FM) because of the strong exchange coupling between sub-lattices, and in contrast to antiferromagnets (AFM), the finite $m_{net}$ enables the read out of FiM magnetic state using tunnel magnetoresistance (TMR) effect.

Recently, the domain wall dynamics near $T_{AMC}$ is predicted to be free from Walker breakdown [9] due to the decoupling of two collective coordinates, and the spin torque is greatly enhanced at the vicinity of $X_{MC}$ [5]. To exploit the rich physics near the two compensation points, one has to unambiguously determine the $T_{MC}$ and $T_{AMC}$. $T_{MC}$ can be determined in many ways, including the direct measurement of magnetization using vibrating sample magnetometer (VSM) [5] or an indirect measurement of the anomalous Hall resistance ($R_{AHE}$) versus magnetic field ($H$) loops at different $T$ with coercivity field ($H_C$) peaks at $T_{MC}$ [2]. However, it remains difficult to experimentally determine the $T_{AMC}$. In addition, the difference between $T_{AMC}$ and $T_{MC}$ can vary from a few K to several tens of K in different samples.

In this Letter, we propose a device structure and show that the $T_{AMC}$ and $T_{MC}$ can be unambiguously determined by conducting two sets of current induced switching. First, we analytically exploit the symmetries in both types of switching. Next, a theoretical model based on the modified Landau-Lifshitz-Bloch (LLB) equation is developed to systematically describe the $T$ dependent FiM dynamics. Finally, numerical simulations are performed to verify our proposal, and the relation between $T_{AMC}$ and $T_{MC}$ in different samples are studied to show the generality of our model.

The device structure, schematically depicted in Fig. 1, consists of a magnetic tunnel junction (MTJ) deposited on top of the HM layer, with the two sets of operations defined in Fig. 1 (a) and (b), respectively. The MTJ includes one perpendicular-FM pinned layer (PL) and one perpendicular-FiM free layer (FL), sandwiched by a spacer layer (e.g. MgO). The FiM in this study is $Gd_X(CoFe)_{1-X}$ where Gd (CoFe) dominates at low (high) $T$. Since CoFe has a larger g-factor, $T_{AMC}$ is higher than $T_{MC}$. In addition, the direction of charge current ($J_C$) is controlled by the voltages (i.e. $V_{A,B,C}$).

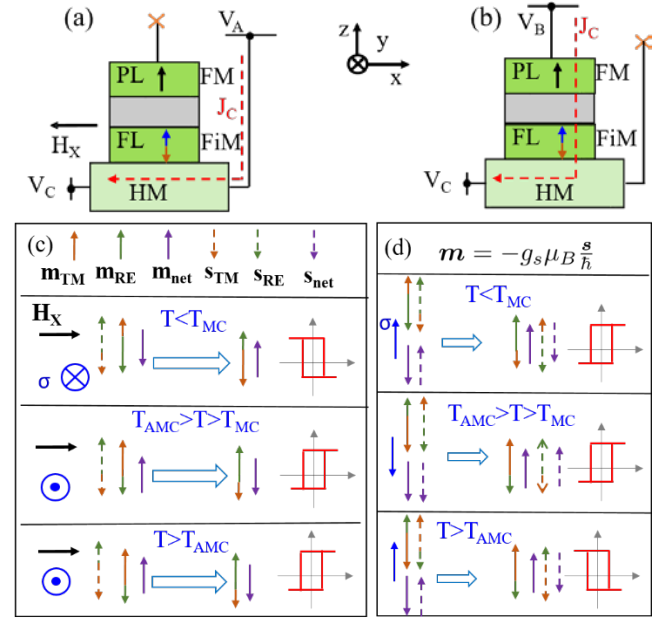

FIG. 1. Device structures with perpendicular FiM-MTJ deposited on the HM layer. The red dash line denotes the charge current, which is controlled by the voltages ($V_{A,B,C}$). The cross symbol denotes the blocking of current path, which can be achieved by using a transistor. (a) The FL is switched by SOT, assisted by the $\boldsymbol{H_X}$. (b) The FL is switched by STT. (c, d) Illustration of $T$ dependent switching corresponding to (a) and (b), respectively.

As shown in Fig. 1 (a), a lateral $J_C$ flowing through the HM layer generates spin orbit torque (SOT) acting on the FiM layer due to the spin Hall effect [10] and Rashba-Edelstein effect [11]. To achieve deterministic switching in the perpendicular direction, an external magnetic field ($H_X$) along the current direction is required [1]. For simplicity, we define it as type-I switching. As we will discuss later, the switching direction in this type is determined by $m_{net}$, and a reversal in switching direction will be observed across $T_{MC}$ by plotting the $T$ dependent $m_{TM}$-$J$ loop. In experiment, $m_{TM}$ can be obtained by measuring $R_{AHE}$ in a Hall bar structure, by noting that $R_{AHE}$ is mainly determined by the magnetic moment of TM element [1]. The switching direction of this type can be understood by analyzing two torques. The first one, $\tau = \Delta m \times H_X$, generated by $H_X$ only determines the switching direction; the other one, $\Delta m = m \times (m \times \sigma)$ where $\sigma$ is the spin polarization, originates from the spin torque and should be sufficient to overcome the energy barrier. It is clear that the switching direction is reversed by using an opposite $H_X$ or $\sigma$, which qualitatively agrees with the experimental result [12]. Fig. 1 (c) illustrates the type-I switching in different $T$ regions. For $T < T_{MC}$, a +y-polarized $\sigma$ generates $\Delta m$ in the −y-direction, resulting in $\tau = +$ z, i.e. $m_{net}$ is switched from down to up. For both $T_{AMC} > T > T_{MC}$ and $T > T_{AMC}$, a −y-polarized $\sigma$ generates $\Delta m$ in the +y-direction, resulting in $\tau = -$ z, i.e. $m_{net}$ is switched from up to down. Therefore, the $m_{TM}$-$J$ loop is reversed across $T_{MC}$.

The operation in Fig. 1 (b) is defined as type-II. The electrons flowing through the MTJ structure are polarized by the PL, and exert spin transfer torque (STT) on the FL by transferring their angular momentum [13, 14]. There is also finite SOT since $J_C$ flows through the HM layer. We neglect it for two reasons: a) it doesn't lead to deterministic switching due to the lack of $H_X$; b) since only the horizontal component of $J_C$ contributes to SOT, the torque is too small to disturb the equilibrium FiM state. As discussed later, the switching symmetry in this type reverses at $T_{AMC}$, which is determined by the nature of spin torque. According to the formula of spin torque ($\tau_{ST} = s \times (s \times \sigma)$), it aligns the angular momentum antiparallel to $\sigma$. For both $T < T_{MC}$ and $T > T_{AMC}$ (see Fig. 1 (d)), $m_{net}$ is opposite to $s_{net}$, and the switching direction follows the conventional spin torque switching [15]. However, the switching pattern is abnormal for $T_{AMC} > T > T_{MC}$, where spin torque aligns $m_{net}$ antiparallel to $\sigma$. This phenomenon was first reported by Jiang et al. [8] in the study of STT switching of CoGd. In this region, $m_{net}$ is parallel to $s_{net}$. Due to the effect of STT, $s_{net}$ is aligned antiparallel to $\sigma$, resulting in an antiparallel configuration between $m_{net}$ and $\sigma$. By plotting the $m_{TM}$-$J$ loop, the switching direction in type-II is reversed across $T_{AMC}$.

To understand these FiM dynamics, a model which can systematically capture $T$ dependence is required. The commonly used Landau-Lifshitz-Gilbert (LLG) model [8, 16-19] is invalid for this purpose since it is based on the fixed magnetization length assumption. To describe the $T$ induced magnetization length change, an additional longitudinal relaxation term is introduced into the LLB model. The model has been widely used to describe the FM dynamics at elevated $T$ [20, 21], and recently the LLB of FiM is also developed to simulate laser-induced switching [22]. In this Letter, we derive the FiM-LLB equation with the spin torque contribution. The dynamics of spin angular momentum at each lattice site is described by the atomistic LLG equation

$$\dot{s} = \gamma[s \times (H + \zeta) - \lambda s \times (s \times H) + H_I(s \times (s \times \hat{z}))], \quad (1)$$

where $H$ is the effective field including anisotropy field and exchange interactions with other atoms, $\zeta$ is the random thermal field, and $H_I$ is the spin torque field. Based on this atomistic equation, the collective behavior of sub-lattices is described by the corresponding Fokker Planck equation

$$\frac{\partial f}{\partial t} + \frac{\partial}{\partial (N)} \{\gamma N \times H - \gamma N \times (N \times (\lambda H + H_I \hat{z})) + \frac{\gamma \lambda T}{\mu_0}[N \times (N \times \frac{\partial}{\partial N})]\} f = 0, \quad (2)$$

where $f$ is the spin distribution function, $N$ is the vector on a sphere with $|N| = 1$, and $\lambda$ is the damping constant. By transforming the spin angular momentum to the magnetization using $m \equiv <s> = \int d^3 N N f(N,t)$ together with the use of mean field approximation [22, 23], the final form of the LLB equation is derived as

$$\dot{m}_v = \gamma_v (m_v \times H_v^{MFA}) - \Gamma_{v,\parallel}(1 - \frac{m_v \cdot m_{0,v}}{m_v^2})m_v - \Gamma_{v,\perp}(\frac{m_v \times (m_v \times m_{0,v})}{m_v^2}), \quad (3)$$

where the subscript $v$ denotes TM or RE element, the $H_v^{MFA} = H_{ext} + H_{A,v} + \frac{J_{0,v}}{\mu_v}m_v + \frac{J_{0,vk}}{\mu_v}m_k$ is the mean field with the exchange coupling coefficient, $J_0$. $\Gamma_\parallel$ and $\Gamma_\perp$ are the longitudinal and transverse damping coefficients, respectively. $m_{0,v}$ is the equilibrium magnetization obtained by solving two coupled Curie-Weiss equations

$$m_{0,v} = B(\xi_{0,v})\frac{\xi_{0,v}}{\xi_{0,v}}, \qquad 4(a)$$

$$m_{0,k} = B(\xi_{0,k})\frac{\xi_{0,k}}{\xi_{0,k}}, \qquad 4(b)$$

with $\xi_{0,v} = \beta\mu_v H_v^{MFA}$.

Eq. 3 contains two coupled equations for TM and RE, which need to be solved simultaneously. Numerical integration of Eq. 3 proceeds using a 4th order predictor-corrector method, i.e. the first 4 steps are obtained by a 4th order Runge-Kutta method, after which, the predictor is calculated using the 4th order multi-step Adams-Bashforth method, and the corrector is computed using a 4th order Adams-Moulton implicit method. The model is verified by benchmarking with experimental $M$-$H$ loop and $M$-$J$ loop [1, 2]. Furthermore, the model can capture the essential physics including the exchange interaction between TM and RE elements, $T$ induced magnetization reduction [23], transition from RE to TM dominant by changing $T$ or $X$ [2, 4], peak of $H_C$ [2] and spin torque [6] at $T_{MC}$. The details of the model derivation and the ability to capture essential physics will be described elsewhere.

Based on the modified LLB model, we then perform numerical simulations to verify our proposals. First of all, the effect of $T$ on magnetization is investigated. As shown in Fig. 2 (a), the magnetizations of TM and RE reduce with $T$ and vanish at a common Curie temperature, $T_C$ = 315 K, which is a unique property of FiM since their FM counterparts have distinct Curie temperatures ($T_{C\_Fe}$ = 1043 K, $T_{C\_Gd}$ = 292 K). This finding also highlights the strong exchange coupling between sub-lattices. The use of different g factors and magnetic moments for TM and RE ($g_{TM}$ = 2.05, $g_{RE}$ = 2, $\mu_{TM}$ = 2.217 $\mu_B$, $\mu_{RE}$ = 7.63$\mu_B$) leads to the separation of $m_{net}$ and $s_{net}$ as shown in Fig. 2 (b). The transition from RE to TM dominant occurs at $T_{MC}$ = 165 K, and the angular momentum transition happens at $T_{AMC}$ = 195 K. In addition, we find that the separation between $T_{MC}$ and $T_{AMC}$ varies from several K to several tens of K, depending on the strength of exchange coupling between sub-lattices. It is worth noting that the results in Fig. 2 are obtained from equilibrium state calculation by solving the coupled Curie-Weiss equations, which cannot be experimentally verified due to the incapability of measuring the angular momentum. As we have discussed, both compensation points can be determined by exploiting the spin torque symmetry around $T_{AMC}$ and $T_{MC}$. The simulation results for $T$ dependent type-I and type-II switching are shown in Fig. 3. In type-I, $m_{TM}$ is switched from up to down under positive current for $T < 165$ K, whereas the switching direction is reversed for higher $T$ (see Fig. 3 (a)). In contrast, the symmetry reversal in type-II occurs at $T = 195$ K (see Fig. 3 (b)). According to the discussion in Fig. 1, $T_{MC}$ and $T_{AMC}$ are determined to be 165 K and 195 K, respectively. The agreement between the equilibrium state calculation (Fig. 2) and the dynamic spin torque switching (Fig. 3) demonstrates the effectiveness of our model and provides an experimentally accessible method to determine $T_{AMC}$.

Finally, we study the relationship between $T_{AMC}$ and $T_{MC}$ in samples with different $X$. As shown in Fig. 4 (a), a good linearity between $T_{AMC}$ and $T_{MC}$ is observed, which can be explained using the $T$ dependent magnetization [24], i.e. the difference in sub-lattices magnetization leads to different compensation points. However, instead of using a simple power-law relation between $M_S$ and $T$ [24], we solve the coupled Curie-Weiss equation to systematically capture the $T$ induced magnetization change. Similar to the atomistic modeling result in ref. [23], we show that both $T_{MC}$ and $T_{AMC}$ only exist in certain region ($0.23 < X < 0.26$). For $X < 0.23$, due to small amounts of RE, the samples are TM dominant for all temperatures, whereas the coincidence of $T_{MC}$ and $T_C$ makes RE dominant for $X > 0.26$. This explains the experimental observation that only certain samples have $T_{MC}$ [1]. The ability to capture $T_{AMC}$, $T_{MC}$ and $T_C$ in different samples demonstrates the generality of our model, and the knowledge of the relation between $T_{AMC}$, $T_{MC}$ and $X$ would be helpful in designing room temperature FiM based electronic devices.

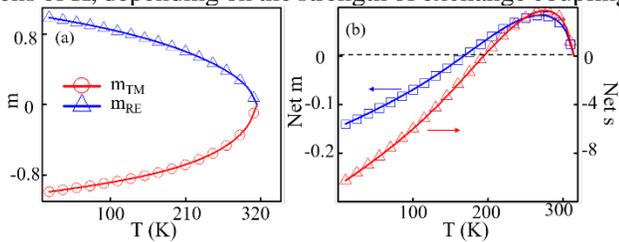

FIG. 2. Effect of $T$ on (a) sub-lattice magnetization, and (b) net magnetization (left y-axis) and net angular momentum (right y-axis). The $m = 1$ is defined at $T = 0$. The intersections of the dash line with the magnetization and angular momentum denote the $T_{MC}$ and $T_{AMC}$, respectively.

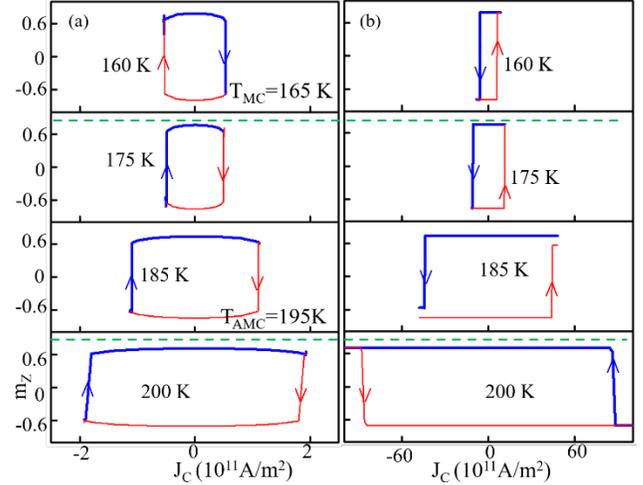

FIG. 3. Magnetization of the TM element versus $J_C$ under different $T$ in (a) type-I, and (b) type-II. The $H_X$ in (a) is 1 mT. The initial magnetization is $m_Z = 1 (-1)$ for the thick blue (thin red) lines. The two dash lines divide the diagram into three regions, i.e. $T < T_{MC}$, $T_{AMC} > T > T_{MC}$, and $T > T_{AMC}$.

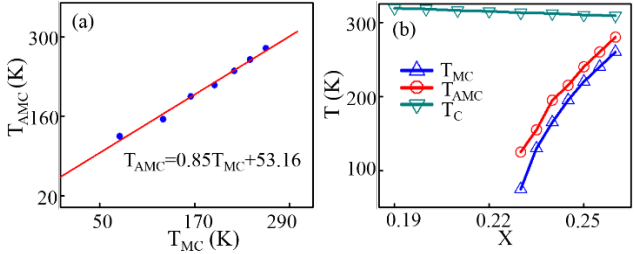

FIG. 4. (a) Relation between $T_{AMC}$ and $T_{MC}$. Blue dots are samples with different $X$ ranging from 0.23 to 0.26, which are fitted using a linear function (red line). (b) The effect of $X$ on $T_C$, $T_{AMC}$, and $T_{MC}$, where $T_{MC}$ and $T_{AMC}$ only exist for $0.23 < X < 0.26$.

In conclusion, we proposed a device structure to identify the $T_{AMC}$ and $T_{MC}$ by conducting $T$ dependent current switching. To systematically describe the FiM dynamics at different $T$, we have extended the LLB model to include the spin torque effect. We verify this proposal by numerically simulating two sets of $T$ dependent switching, and the results agree with the equilibrium state calculation using Curie-Weiss equations. Furthermore, the recent reported linear relation between $T_{AMC}$ and $T_{MC}$ can be well explained using our model. The work should facilitate the study of current excited dynamics in FiM based devices.

We acknowledge financial support from CRP award no. NRF-CRP12-2013-01 and MOE2013-T2-2-125.

a)a0132576@u.nus.edu, b)elelg@nus.edu.sg